\documentclass[10pt]{article}

\usepackage{times}
\usepackage{amsfonts}
\usepackage{amssymb}
\usepackage{amsbsy}
\usepackage{geometry}
\usepackage[small,sf,centerlast]{caption} 
\usepackage{graphicx}
\usepackage{floatflt,epsfig}

\def\Mbol{M_{\rm bol}}
\def\teff{T_{\rm eff}}
\def\mast{M_\ast}

\def\msol{M_\odot}

\def\last{L_\ast}
\def\mast{M_\ast}

\def\llsol{L/L_\odot}

\def\**{\par\hangindent=2.0cm\hangafter=1} 

\begin{document}

\begin{Large}
\centerline{\textsc{On the Robustness of the Cepheids' P-L relation}}
\end{Large}
\medskip
\centerline{Alfred Gautschy, Astronomisches Institut der Universit\"at Basel}
\smallskip	
\centerline{Venusstrasse 7, 4102 Binningen, Switzerland}
\bigskip

\hrule\smallskip 
Based on a talk delivered at the Minisymposium
``Recent Results on H$_0$'' during the 19$^{\rm th}$ Texas Symposium
on Relativistic Astrophysics, Paris.
\smallskip\smallskip\hrule

\section{Introduction}

The Cepheids' period~--~luminosity (P-L) relation got itself talked
about again in the recent past due to claims of observed metallicity
dependences of at least its zero point (Gould 1994, Sasselov et
al. 1997, Sekiguchi \& Fukugita 1998, Kennicutt et al. 1998).  The
divergences of the observational results obtained by the different
authors are large, however. Therefore, to date the situation
concerning the magnitude and the reliability of the effect is rather
uncertain on the observational side. Along the theoretical avenue,
some progress towards more realistic and in particular more consistent
Cepheid modeling has been achieved recently. Upper limits on the
reaction of the P-L relation on assumptions in evolution computations,
pulsation stability analyses, and mappings onto photometric passbands
can be estimated now quite reliably.

This informal review discusses some of the progress (relying mostly
on an extensive comparison by Sandage et al. 1998) and it emphasizes
modeling problems which still have to be overcome to finalize the
theoretical foundation of the Cepheids' P-L relation.

\section{Stellar Evolution}
In absence of proper models for stellar hydrodynamics in the framework
of stellar evolution, astronomers are urged to parameterize processes
such as convective overshooting and semiconvection at various levels
of accuracy.  Numerous free parameters always go along with such
modelling attempts. As different schools tend to favor different
choices of such model parameters, we encounter frequently hard to
follow controversies and dichotomies in the literature.  For a
nonspecialist it becomes then essentially impossible to select a
particular simulation result for his purpose based on rational
reasoning alone. As recipe-driven modeling continues, this unfortunate
situation will hardly disappear in the near future.

Semiconvection and convective overshooting are emphasized above
because they are the hydrodynamical key-phenomena whose effects cause
major uncertainties in the stellar evolution of intermediate-mass and
massive stars. Evidently, any other modification of the stellar
microphysics shifts and twists the star's track on the
Hertzsprung-Russell (HR) plane too. In contrast to hydrodynamical
processes, however, broad acceptance of say opacity tables and
nuclear-reaction rates is found in the community.

Numerical observations reveal that overshooting shifts the
evolutionary track of a given stellar mass mainly to higher luminosity
during the main sequence phase. During the core helium burning the
size of the blueward loops is modified when overshooting is
incorporated in the numerical scheme (e.g. Matraka et al. 1982,
Stothers \& Chin 1992, Alongi et al. 1993). In contrast to
overshooting, the effect of semiconvection appears to be more
dramatic, at least during the core-helium burning phase.  Even the
very existence of blue loops can depend on incorporating
semiconvection (e.g. Langer et al. 1985, Langer 1991).

Even if the physical processes and the micro-physical data are the
same in different evolution codes, the resulting tracks can differ
significantly.  A comparison of Baraffe et al.'s (1998) Fig.~1 with
Saio \& Gautschy's (1998) (SG98) Fig.~1 exposes that in the first case
an additional smaller loop occurs superposed on the the larger (main)
blue loop during the core helium burning. Hence, where SG98 find only
three crossings of the Cepheids instability domain, Baraffe et
al. (1998) can accommodate up to five.
  
The argument that the second crossing of the instability region only
is relevant for the Cepheid phenomenon is perpetuated even in the most
recent literature. A closer look at stellar evolution data shows that
only $M_\ast \lessapprox 6 M_\odot$ spend by far the longest time in
the instability domain during the second crossing (say the ratio of
the residence time therein exceeds 10 compared with the other
crossings). Towards higher masses, the ratios of 2nd to 3rd and 2nd to
1st crossing times drops markedly and therefore the probability to
encounter 1st and 3rd time crossing stars increases to a
non-negligible level.

\begin{figure}
  \begin{center}
  \includegraphics[width=7cm]{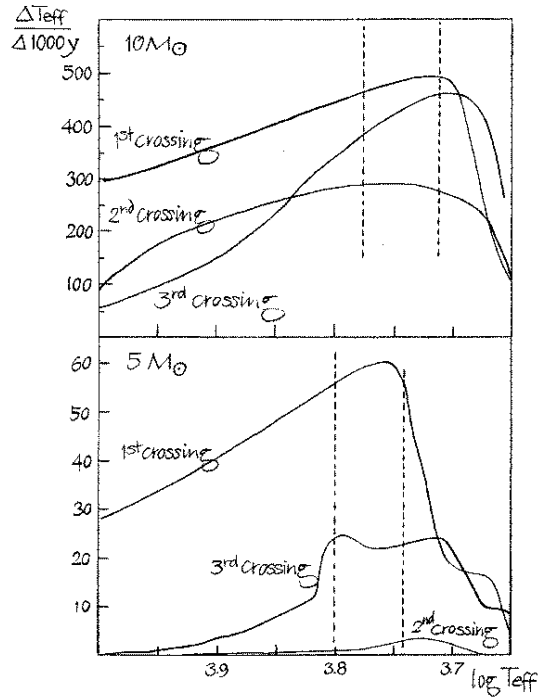}
  \end{center}
     \caption{Evolutionary effective temperature changes per 
         millennium of 5 and 10 $\msol$ models crossing the instability strip 
         region. Its approximate boundaries are indicated by dashed lines. 
         The heavy element abundance  $Z$ is 0.004.
	     }
  \end{figure}

For a $7 \msol$ star, the ratio of 2nd to 1st crossing time reduces to
about 3 and to 1.6 for 3rd/2nd crossing. The numbers further decrease
as the stellar mass increases.  Detailed numbers depend sensitively on
the choice of the abundances.  Saio (1998) presented a figure
showing that for his computations of $X = 0.7, Z=0.02$ sequences the
3rd crossing constitutes the slowest passage through the strip for
stellar masses above $5 \msol$.

All the uncertainties and the resulting discrepancies in the
evolutionary tracks as well as, of course, any changes of the star's
chemical composition modify the mass~--~luminosity (M-L) and the
mass~--~radius (M-R) relation at the the position of say the
fundamental blue edge (FBE) of the instability strip.  How this
translates into and influences the P-L relation will be addressed in
the next section.

\section{Stellar Pulsation Theory}
The basic physical mechanism underlying the pulsations of Cepheids is
well understood. This success story started in early sixties
(cf. Baker \& Kippenhahn 1965 and Cox 1980 for further references).
This does not mean, however, that all theoretical problems have been
resolved in full detail. The qualitative era has passed and
increasingly more accurate observational data need to be explained
with a theoretical framework in which stellar evolution and also
micro-physics is known and understood at a high level.

To motivate the existence of a P-L relation for Cepheids we start with
the pulsation equation:
$\Pi\sqrt{\langle{\rho_\ast}\rangle/\langle{\rho_\odot}\rangle} = Q $,
with $\Pi$ being the theoretical period (in contrast to the observed
one $P$). Under favorable circumstances, i.e. under appropriate
relations between stellar mass, $\mast$, its luminosity, $\last$, and
the effective temperature, $\teff$, a simple relation between period
and luminosity emerges. Note that the pulsation constant, $Q$, is
constant only for homologous models.  Theoretically, the formulation
of the P-L relation at the position of the blue edge (in the case of
Cepheids at the fundamental blue edge:FBE) is the least cumbersome one
and therefore it is chosen in the following. Observers, on the other
hand, prefer something like the \emph{ridge line}, which is a mean
line through the observational data on the P-L plane. A ridge line is
statistically much easier to compute than defining an accurate
envelope on top of sparse data.  If theorists would comply with the
observers' choice they would have to have a clear idea about the width
of the instability strip, which is, however, hard to come by
computationally. The position of the red edge of the strip is defined
by the efficiency of the convective leakage of the flux in the
superficial stellar layers~--~an extremely cumbersome task to model!
We return to the role of convection further below.

After some easy manipulations we can write the pulsation equation as
\begin{equation}
 \log \Pi = - {1 \over 2} \log M / M_\odot
            + {3 \over 4} \log L / L_\odot
            - 3           \log \teff
            +             \log Q             + C
\end{equation}

In the first term of the rhs we can assume a M-L relation at the
position of the FBE, i.e. $M = f_1(L;q_i)$. Notice that this relation
depends on all those assumptions (indicated by the $q_i$'s) which
influence the loci of the tracks around the instability strip. These
quantities are, however, often buried deep in the formulation of the
modeling of stellar evolution and therefore they are difficult to
quantify.  Furthermore, the M-L relation depends on the evolutionary
phase, or in other words, in depends on the number of crossing of the
instability strip.

The third term on the rhs of eq.~(1) can be eliminated via a
parameterization of the position of the FBE, i.e. $\teff = f_2(L,q'_i)$
which depends implicitly again on various quantities $q'_i$(which might
be different from the ones influencing the M-L relation). 

Essentially, the quarreling about the robustness of the P-L relation,
i.e. the uniqueness of its slope and zero-point, after $f_1, f_2$ are
introduced into eq.~(1), boils down to quantifying the effect of the
$q_i, q'_i$ on the P-L relation. From first principles, little is to
be learned on these matters. Therefore, numerical calculations have to
be performed and conclusions (even if not so conclusive) have to be
derived from such ``numerical observations''.

Coarsely speaking, the M-L relation, the parameterization of the FBE,
and the M-R relation of supergiants in the Cepheid stage all show
significant dependences on $Y$ and $Z$ abundances, on the number of
the crossing (i.e. the evolutionary stage) of the instability region
(cf. SBT98), and on all the stellar hydrodynamical uncertainties
(e.g. mixing length, overshooting length-scale, semi-convective
efficiency).  Most surprisingly, however, the \emph{combined} effect
of the various dependencies leads eventually to a nearly $Y, Z$, and
crossing-number independent P-L relation.

As an example consider the following: The FBE (at any luminosity)
depends on $Y$ and $Z$ roughly as
\begin{equation}
	\Delta \log \teff = +0.04\,\Delta Y - 0.49\,\Delta Z 
\end{equation}
(SG98). More extensive (and therefore probably more
reliable) studies of the Y dependence show a larger coefficient associated
with $\Delta Y$: +0.14 in Chiosi et al. (1993), and +0.11 in Iben \&
Tuggle (1972). The important thing to notice is the counteracting
influence of $Y$ and $Z$. It reduces the shift of the FBE when Z is
modified with the frequently encountered cosmogonical constraint of
$\Delta Y / \Delta Z \approx 3 - 5$.

Furthermore, despite the varying M-L relation for different heavy
element abundances and different hydrodynamical assumptions concerning
convection, the M-R relation adapts such that finally the
slope of the emerging P-L relation changes by less than about 10 \%.
The same applies for the different crossing numbers across the
instability region: The M-L relation is a different one for each
crossing, so is the M-R relation. In combination they result in a
zero point shift of less than $0.1 \Mbol$ for any choice of chemical
abundances. The slope remains unchanged to a high degree.   

Summa summarum: The combination of stellar evolution and linear
stability analyses on these models directly leads to a P-L relation at
the FBE (at least) which is not noticeably dependent neither on
chemical abundances nor on hydrodynamical treatment or evolutionary
stage.  The average scatter of the P-L relation at the FBE is about
0.047 in $\log \llsol$. The ad hoc defined fundamental red line (FRL)
of (SG98)~--~postulating the fundamental red edge
(FRE) to be parallel-shifted in $\log \teff$ by 0.06 indicates a P-L
relation for the FRL running also parallel to the one of the FBE; this
result is not a priori trivial as the stars evolve non-homologously
across the instability strip and as the efficiency of convective leakage
might be a strong function of $\teff$.

Most of all, however, one of the major unsettled issues in stellar
pulsation theory~--~ the influence of convection~--~is important also
for Cepheids. Modeling of convection in pulsating stars has been
attempted with mixed success for many years (e.g. Unno 1967, Baker \&
Gough 1979, Stellingwerf 1982, Xiong et al. 1998, Yecko et
al. 1998). Usually it is multiply parameterized and therefore it is
not satisfactorily understood.  Model computations clearly demonstrate
that convective leakage is the dominating process to cause a red edge
and therefore it is an indispensable ingredient of the theory.  A
proper treatment of convection might also be necessary to pin down the
blue edge with high accuracy. The sensitivity of blue edge on
convection depends on the stellar mass, however.  As the instability
strip is tilted to lower effective temperatures for higher
luminosities, higher-mass stars will be influenced stronger by
convection than low-mass stars (cf. Fig.~2).

At high luminosities, radiation pressure tends to destabilize stellar
envelopes (i.e. tends to shift the FBE to the blue) whereas convective
leakage in the superficial layers induces a redshift of the FBE. The
competition between these two effects will determine the final
outcome. Taking the observational P-L data at face value, it appears
as if convection is the dominating effect, i.e. the shift of the FBE
to the red induces a reduction of the slope of the P-L relation at its
upper end (cf. SBT98). The deviation from simple
linearity of the P-L relation appears to set in above about 60 days
period. This is relevant for stellar masses above 10 $\msol$. 

\begin{figure}
  \begin{center}
   \includegraphics[width=8cm]{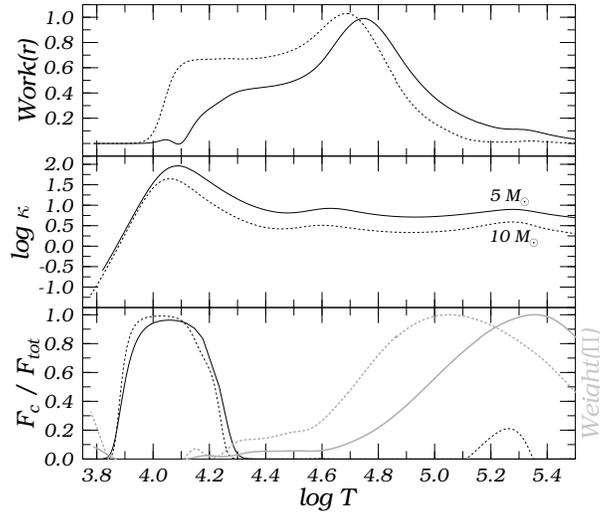}
  \end{center}
    \caption{Radial structure of a 5 (full lines) and 10 $\msol$ 
	(dashed lines) Cepheid
	envelope at the FBE. The top panel shows the total
	workintegral with the positive contributions in the partial
	ionization zones of H/HeI and HeII. The middle panel shows the
        Rosseland opacity structure with the appropriate bumps. The
	bottom panel displays the fraction of the energy 
	flux which is transported by  convection in the H/HeI
	ionization zone (and some convective flux in the Z-bump region
        of the 10 $\msol$ model. In grey we show the arbitrarily normalized 
        adiabatic weight functions for the overstable fundamental modes.
            }
  \end{figure}

The lowest panel of Fig.~2 shows that for the 5 and the 10 $\msol$
stars convective transport is the dominating energy transport
mechanism within the partial H/HeI ionization zone: locally more than
90 \% of the flux is transported by material motion. The integral
work, which is shown in the top panel, is not significantly overlapping
with the convection-dominated region in the 5 $\msol$ model. Most of
the driving happens deeper inside the envelope, in the HeII partial
ionization zone around $\log T = 4.75$. The situation changes,
however, at 10 $\msol$: More than 50 \% of the driving occurs there
already in the H/HeII region which overlaps with about half of the
convective flux dominated region of the envelope. For the latter
model, a neglect of the convective flux perturbation in the stability
analysis will most probably affect the eigenanalysis markedly. In
other words, a simple analysis, suppressing convective leakage even at
the blue edge will produce too blue a FBE compared with a realistic
treatment. Therefore, over the whole period range, say up to 100 days,
a simplistic computation will produce a P-L relation with a slope
which is larger than what nature realizes. When constraining a P-L
parameterization to low luminosities (i.e. shorter periods), the
effect of a even a suppressed pulsation-convection coupling might lead
to a slight zero-point shift in the P-L relation. Only a slope shift
is not so likely.  The magnitude of such a shift is hard to predict.
A comparison relies on how realistically convection penetrating
optically thin regions can be described. Presently, we are far from a
physically realistic treatment of this domain.

In contrast to repeated mis-citations (e.g. Bono et al. 1998) SG98 did
not use ``radiative'' models in their linear, nonadiabatic stability
analyses. The used evolutionary stellar models which are convective
wherever stellar thermodynamics requires it. What is not included,
however, is the \emph{perturbation} of the convective flux due to some
convection-pulsation interaction prescription. This clarification
seems important as the correct (radiative/convective) structure of the
equilibrium model is relevant for the nonadiabatic \emph{period} derived
from a linear analysis~--~even if the perturbation of the convective
flux turns out to be irrelevant for the onset of pulsational instability for
the short-period Cepheids. Of course, the same aspect pops up again
when comparing linear models with convective non-linear ones (e.g. BMS
1999).

The lowest panel of Fig.~2 shows in grey the weight functions of the
fundamental modes for the $5 \msol$ and the $10 \msol$ models. The
magnitude of the arbitrarily normalized weight function measures the
local importance of the stellar properties to determine the period the
pulsation mode under consideration. Obviously, the convection zone in
the H/HeI partial ionization region has marginal influence on the
weight function for both stellar masses. It is clearly seen that the
two sets of curves live in essentially disjoint regions of the
star. However, it is important to note for comparisons it is relevant
to compute stellar structures with resulting from the same physical
input: If the convection zones are suppressed to compute radiative
models (e.g. BMS 1999) then the compressibility structure ($\rho(r) -
T(r)$) differs from the one in a radiatively/convectively layered
models. Very naturally, two such sets of such unlike models will have
different distributions of weight functions and therefore different
periods. The more extended the convection zones become, the more the
periods diverge between the two approaches: This is just what is seen
in Fig.~53 of BMS99.
 
Modern nonlinear modeling which includes diffusion-type convection
treatment coupled with the star's envelope pulsation do exist
(cf. BMS99). The convection formulated therein is a refined version of
Stellingwerf's (1982) approach. The BMS99 nonlinear FBE is at all
luminosities below $10\,000$ $\llsol$ about 200~K hotter than e.g. SG98 FBE
without pulsation-convection coupling. Na{\"\i}vely, we would have expected
the opposite behaviour. Above $10\,000$ $\llsol$ the BMS99 FBE shifts
coolward rather abruptly, suggesting a FBE at 5500~K at $32\,000$
$\llsol$ where the extrapolation of the SG98 data leads to 5700~K.
This drop in temperature of the FBE is the reason of the quadratic
term introduced into the P-L parameterization by Bono \& Marconi
(1998). Observations from LMC and SMC do not, in our opinion, support
such an efficient convective leakage at long periods (high
luminosities).

Considerable complications enter the treatment of convection in
Cepheids since the superficial convection zone associated with the
H/HeI partial ionization reached into optically thin regions. Hence,
radiative losses become important in the energy balance of the
convective elements. This is very difficult to model; to date a
reliable quantitative description is not available. For the finite
amplitude behaviour, i.e. for the lightcurve, the details of the
treatment of convection at the outer boundary was shown to be
important already in RR Lyrae stars (Feuchtinger 1998) which are
hotter than the Cepheids.

It is unlikely that the final effect of convection on the P-L relation
will introduce a metallicity dependence which we are missing
currently.  The driving in the outermost regions is H-ionization
dominated, and this is very robust against envisaged small changes in
the chemical composition.

\begin{figure}
  \begin{center} 
    \includegraphics[width=10cm]{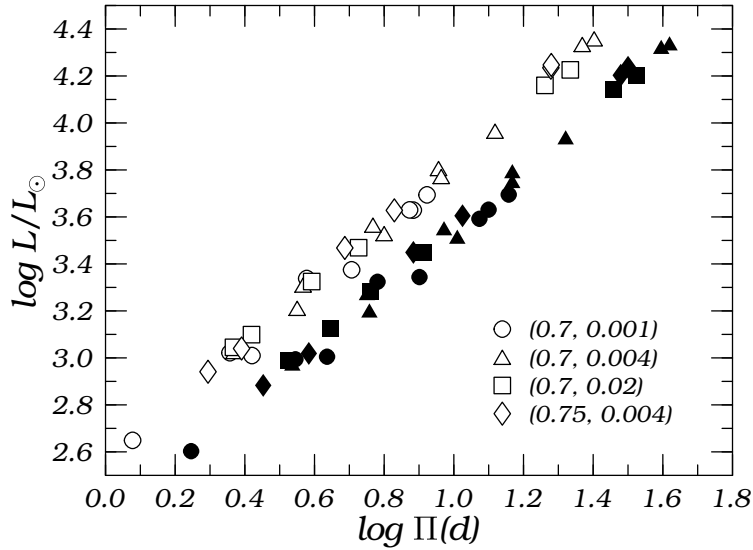} 
  \end{center}
    \caption{The P-L relation according to SG98. FBE data for different
    crossings and various abundances are shown as open symbols.  Full
    symbols show the locus of an ad hoc defined red line (offset by
    $\Delta \log \teff = 0.06$ from the FBE) on the P-L plane.  }
\end{figure}

To complete this section, we write the P-L relation as shown in
Fig.~3, as $\Mbol = a \log \Pi + b$. A comparison of several
theoretical studies by SBT98 demonstrated that at 10 days period, the
zero point of th P-L relation agrees to within $0.1 \Mbol$ for the
different authors (and therefore different approaches in terms of
stellar evolution and stellar stability computations).  With the SG98
data we arrive at $\Mbol = -4.84$ (at [Fe/H] = 0) or at $M_V = -4.92$
after bolometric correction according to SBT98.  The zero-point
remains stable at a level of about +/- 0.05 mag when [Fe/H] varies
between +0.4 and -1.7 and when Y changes (independently) between 0.25
and 0.30. The agreement is somewhat degraded at $\log P = 1.5$ since
the slopes derived from the different sources vary at the level of
about 10 \%. In the study of SG98 the value of $a$ was found to
increase by about 9 \% upon reducing [Fe/H] from zero by one dex.
This all shows nevertheless the rather remarkable stability of the P-L
relation against changes in physical assumptions, numerical
realizations, and last but not least against abundance variations.

\section{Stellar Atmospheres}

Even if we should have gained some confidence in the computations of
stellar evolution and then in stellar stability properties,
there is another hurdle to clear: the transfer of the modulated energy
flux through the stellar atmosphere. 

As we have seen in the previous sections, the \emph{theoretical}
Cepheid P-L relation appears not to be very sensitive (compared with
the level of accuracy of presently obtained observations) on
metallicity and assumptions on the microphysics or on hydrodynamical
processes. Even if that should prove correct in the future, the
transformation into observationally relevant filter passbands can
potentially destroy this simplicity due to differential blanketing
effects in the stellar atmosphere.

Several recent studies of the P-L relation performed a mapping of
bolometric magnitudes into various photometric passbands (Baraffe et
al. 1998, Bono et al. 1998, SBT98). Only the SBT98 study uses stellar
atmospheres which are constructed for the particular needs of
Cepheids. In the other studies, the available $\log g$ range of the
atmosphere models is too narrow for Cepheids so that extrapolations
are necessary for colors and bolometric corrections.

SBT98 find a strong dependence of color indices (in particular at
short-wavelength) and bolometric correction on metallicity and
gravity. The superposition of these dependencies, when computing the the
P-M$_x$ relations (where $x$ stands for some filter passband), conspires
again to result in a remarkable smallness of the effect when
metallicity is varied by a factor of about 50. 
Table~1 quantifies the metallicity dependence as deduced by SBT98.

 \begin{table}
    \caption{dM$_x$/d[Fe/H]}
      \begin{center}  
	\begin{tabular}{ccc}
        \hline                               \\[-2.5 mm]
	Filter  &         10d   &     31.6d  \\
	\hline\hline                         \\[-2.5 mm]
	B       &       +0.02   &    +0.08   \\
	V       &       -0.08   &    -0.08   \\
        I       &       -0.10   &    -0.10   \\
	\hline
	\end{tabular}
      \end{center}
 \end{table}

The typical scatter in the data amounts to about 0.02 mag. The
variations of the absolute magnitudes at different periods is close to
the noise in the theoretical models, but also in the range of the
uncertainty of observational data. In any case, the above numbers are
more than a factor three smaller than some observational results
published recently (e.g. Gould 1994, Sasselov et al. 1997, Sekiguchi
\& Fukugita 1998, Kennicutt et al. 1998).

Applying the above results to distance moduli obtained in different
passbands for LMC and SMC very good internal agreement is found,
usually on the level of +/- 0.02 mag when using either the SBT98 or
the DiBenedetto (1997) data. Additionally, SBT98 demonstrated that
distance moduli to the Clouds derived with RR Lyr stars and Cepheids
agree well even when not correcting the Cepheids' moduli for
metallicity. This latter result already hints at only a weak
dependence of the P-L zero points on abundance.

\section{Final comments}

Based on the mapping of bolometric relations onto selected filter
passbands with appropriate stellar (static, plane-parallel) atmosphere
models, SBT98 found that a weak metallicity-dependent zero point shift
in the P-L relation exists. Its magnitude is below 0.1 mag/dex of
[Fe/H] in the passbands of B,V, and I. An more exact number could not
be deduced to date as the results scrape close to the scatter brought
about by the models themselves and the different modeling assumptions.
It appears, however, that the theoretically deduced variation with
metallicity lies in the range of the claimed observational
errors. Therefore, from the theoretical side, no significant zero
point variation is expected to be found in the observations presented
to date.  Applying to the distance moduli corrections to the absolute
magnitudes assuming the upper limits of their [Fe/H] dependence leads
to the distance moduli with a remarkable internal agreement in B,V,
and I for both LMC and SMC (cf. SBT98).  The obviously much stronger
[Fe/H] dependences claimed in the observational literature is not
supported by theory and the discrepancy has clearly to be resolved in
the near future as it poisons the use of pulsators to accurately
calibrate the distance scale in the neighboring universe.

One of the important questions at the interface of observations and
theory is that of proper averaging observational data. Theory always
discusses equilibrium quantities and observations provide at best
\emph{some} mean values.  How well do these mean values represent
equilibrium quantities?  The transformations of observational data
averaged over a cycle, such as e.g. $\langle m_{\rm V} \rangle$, to
obtain quantities predicted by theory (such as $M_{V}$ or $\Mbol$) are
unfortunately not unique.  For a brightness to luminosity
transformation, a cycle average of the \emph{intensity} is expected to
be the best as it can be physically motivated. Surprisingly enough,
Karp (1975) shows for his model that a magnitude mean approximates the
equilibrium value best. For colors the situation becomes even more
inscrutable as the quality of a particular choice of averaging is
furthermore considerably passband dependent. Also Bono et al. (1998)
point out these differences without suggesting, however, a
approximation strategy for observers yet.

Another aspect which might lead to unexpected divergences when
defining some sort of a ridge line is the uneven population of the
instability strip.  Taking Fernie's (1990) or SBT98's results at face
value one observes that the Cepheids tend to crowd the blue side and
avoid the red part of the instability strip altogether at short
periods. From the view-point of stellar evolution theory this is not
understandable. The stars should enter the instability strip from the
red and therefore populate the strip from cool to blue upon increasing
the luminosity. Seemingly the opposite is observed!  Furthermore, our
Fig.~1 shows that the evolutionary timescales are not even across the
strip. Almost generically, the evolution is slower close to the blue
edge than around the red one. This seems to be true independently of
the crossing number. It is the second crossing of the 10 $\msol$ star
only which suggests an even speed and therefore an even population of
variables across the instability region.  Not accounting for the above
mentioned population-density differences of pulsators across the
strip, not to mention yet observational biases due to an
amplitude~--~effective temperature~--~period dependence, distorts the
slope of the ridge line. For example, computing a ridge line assuming
a uniform population of the strip and neglecting the short-period blue
clumping discussed by Fernie (1990) leads to a shallower P-L relation
than what one obtains from a FBE relation.

On theoretical grounds, using a FBE relation would be the most
favorable choice. This, however, is hard if not impossible at all to
realize considering the small number of stars usually accessible. When
using  \emph{all} the Cepheids in the strip to establish a slope of
the P-L relation a possible uneven population of the instability strip
should be accounted for. A quantification of the effect should be
feasible for stellar evolution simulators as they have the
necessary data at hand.

\bigskip\bigskip
\noindent 
Acknowledgements: The Swiss National Science Foundation supported this
study through a PROFIL2 fellowship. I am indebted to H. Harzenmoser
for inquisitive discourses during a Meringues-meeting at Chemmeribode
Bad.

\bigskip\bigskip

\centerline{\textbf{References}}
\**
Alongi M., Bertelli G., Bressan A., 
        Chiosi C., Fagotto F., et al. 1993,
	A\&AS 97, 851
\**
Baker N.H., Gough D.O. 1979, 
	ApJ 234, 232
\**
Baker N.H., Kippenhahn R. 1965, 
	ApJ 142, 868
\**
Baraffe I., Alibert Y., M\'era D., Chabrier G., Beaulieu J.-P. 1998
	ApJ 499, L205
\**
Bono G., Marconi M. 1998, in 
	IAU Symp. 190, New Views of the Magellanic Clouds, 
        eds. Y.-H. Chu, J. Hesser \& N. Suntzeff, ASP Press 
\**
Bono G., Marconi M., Stellingwerf R.F. 1999, 
        to appear in ApJS (BMS99)
\**
Bono G., Caputo F., Castellani V., Marconi M. 1998,
	astro-ph/9809127
\**
Chiosi C., Wood P.R., Capitanio, N. 1993,
	ApJS 86, 541
\**
Cox J.P. 1980,
	Stellar Pulsation, Princeton:Princeton University Press
\**
DiBenedetto G.P. 1997,
	ApJ 486, 60
\**
Fernie J.D. 1990, 
	ApJ 354, 295
\**
Feuchtinger M. 1998,
	A\&A 337, L29
\**
Gould A. 1994, 
	ApJ 426, 542
\**
Iben I., Jr., Tuggle R.S. 1972,
	ApJ 178, 445
\**
Karp A.H. 1975,
	ApJ 200, 354
\**
Kennicutt R.C., et al. 1998, 
	ApJ 498, 181
\**
Kochaneck C.S. 1997,
	ApJ 491, 13
\**
Langer N. 1991, 
	A\&A 252, 669
\**
Langer N., El Eid M.F., Fricke K.J. 1985,
	A\&A 145, 179
\**
Matraka B., Wassermann C., Weigert A. 1982,
	A\&A 107, 283
\**
Saio H. 1998, 
	in Pulsating Stars, eds. M. Takeuti and D.D. Sasselov, 
       	Tokyo:Universal Academy Press
\**
Saio H., Gautschy A. 1998, 
	ApJ 498, 360 (SG98)
\**
Sandage A., Bell R.A., Tripicco M.J.  1998,
	preprint
\**
Sasselov D.D., et al. 1997, 
	A\&A 324, 471
\**
Sekiguchi M., Fukugita M. 1998,
	Observatory 118, 73
\**
Stothers R.B., Chin C-W. 1992, 
	ApJ 390, 136
\**
Stellingwerf R.F. 1982,
	ApJ 262, 330
\**
Unno W. 1967,
	PASJ 19,140
\** 
Xiong D.R., Cheng Q.L., Deng L. 1998,
	ApJ 500, 449
\**
Yecko P.A., Koll\'ath Z., Buchler J.R. 1998,
	A\&A 336, 553
\**
\end{document}